# Topological insulators vs. topological Dirac semimetals in honeycomb compounds

Xiuwen Zhang[§,†,‡,*], Qihang Liu[†,‡], Qiunan Xu[♮,‡], Xi Dai[#], and Alex Zunger[†,*]

[§]Shenzhen Key Laboratory of Flexible Memory Materials and Devices, College of Electronic Science and Technology, Shenzhen University, Guangdong 518060, China

[†]Renewable and Sustainable Energy Institute, University of Colorado, Boulder, Colorado 80309, USA

[♮]Beijing National Laboratory for Condensed Matter Physics, and Institute of Physics, Chinese Academy of Sciences, Beijing 100190, China

[#]Department of Physics, The Hong Kong University of Science and Technology, Hong Kong

**ABSTRACT:** Intriguing physical property of materials stems from their chemical constituent whereas the connection between them is often not clear. Here, we uncover a general chemical classification for the two quantum phases in the honeycomb ABX structure—topological insulator (TI) and topological Dirac semimetal (TDSM). First, we find among the 816 (existing as well as hypothetical) calculated compounds, 160 TI's (none were noted before), 96 TDSM's, 282 normal insulators (NI's), and 278 metals. Second, based on this classification, we have distilled a simple chemical regularity based on compound formulae for the selectivity between TI and TDSM: The ABX compounds that are TDSM have B atoms (part of the BX honeycomb layers) that come from the Periodic Table columns XI (Cu, Ag, Au) or XII (Zn, Cd, Hg), or Mg (group II), whereas the ABX compounds whose B atoms come from columns I (Li, Na, K, Rb, Cs) or II (Ca, Sr, Ba) are TI's. Third, focusing on the ABX Bismide compounds that are thermodynamically stable, we find a structural motif that delivers topological insulation and stability at the same time. This study opens the way to simultaneously design new topological materials based on the compositional rules indicated here.

## I. Introduction

Topological insulators (TI's)[1-4] and topological Dirac semimetals (TDSM's)[5,6] are two classes of quantum phase, both having in their bulk band structures an inverted order of the occupied valence and unoccupied conduction bands at the time-reversal invariant momenta (TRIM) in the Brillouin zone (BZ). In three-dimensional TI compounds (e.g., $Bi_2Se_3$ [7]), such bulk band inversion is characterized by a topological invariant[3] $Z_2 = 1$ and leads to the appearance at the bulk-terminated surface of linearly dispersed ('massless') and mutually crossing ('metallic') energy bands, a construction referred to as surface Dirac cones.[8] In three-dimensional TDSM compounds (e.g., $Na_3Bi$ [5]) on the other hand, band inversion is characterized by a two-dimensional topological invariant $v_{2D} = 1$,[9] and leads to the appearance of Dirac cones already in the bulk, a construction protected by crystalline symmetries.[5] Note that TDSM is distinct from non-topological DSM such as the hypothetical $BiO_2$ compound in the cristobalite structure, where a topological invariant cannot be defined.[10] The existence of a given compound as TI or TDSM depends sensitively on structure and composition and is a rather consequential distinction because these two quantum phases have very different physical properties (e.g. quantum spin Hall effect[4] in CdTe/HgTe quantum well TI versus giant magnetoresistance[11] in $Na_3Bi$ TDSM) and different potential applications. However, one is hard pressed to guess based on structure and composition alone which compound will be non-topological and which will be topological, and for the latter, which will be a TI and which will be TDSM. We address this question via direct calculations of the topological invariants for many compositions within the context of a given structure—the honeycomb lattice—and uncover intriguing *chemical regularities*.

The ABX honeycomb structure (ZrBeSi-type, space group $P6_3/mmc$; No. 194) consists of BX planar honeycomb layers and "stuffing" A layers in between these layers. Our interest in the honeycomb lattice stems from a few reasons: (i) a closer inspection (see below) of the predicted TI's in groups I-XI-VI (such as LiAuTe[12]) and I-XII-V (such as LiHgSb[12]) honeycomb compounds reveals that they are in fact all TDSM's. Other previously assigned topological materials in group XI-II-V (such as AgBaBi[13,14]) are also established to be TDSM's. *Thus, at this point there are no material realizations of TI in ABX honeycomb compounds*. (ii) The special crystalline symmetries in the honeycomb ABX structure offer the opportunity to realize a variety of interesting physical properties even beyond TI and TDSM, such as hidden spin polarization[15] and hourglass Fermion[16]. (iii) There are ~800 (real as well as hypothetical) compounds in this group, providing a diverse range of chemical bonding, so the conclusions drawn are likely to be rather general. In this study, we performed high-throughput calculation of the ~800 octet ABX compounds (see Figure S1 in Supporting Information). Whereas most of the ABX compounds in Fig. S1 do not adopt the honeycomb structure (compounds indicated in Fig. S1 in red color are stable in the honeycomb structure, others have different structures), we have considered in the first step of this calculation all com-pounds in the honeycomb structure so as to establish the particular structural motif (Section II.B) that makes some compounds TI, others TDSM. In the second step (Section II.C) we establish for a subset of compounds that are the most interesting, the lowest-energy structure. From the above calculations, a simple chemical regularity based on compound formulae for the selectivity between TI and TDSM is distilled. This opens the way to design topological materials with intriguing physical properties from the basis of chemical constituents. It also shed light on the correlation between topological properties and local structural motifs, suggesting design of TI's using local motifs as building blocks.

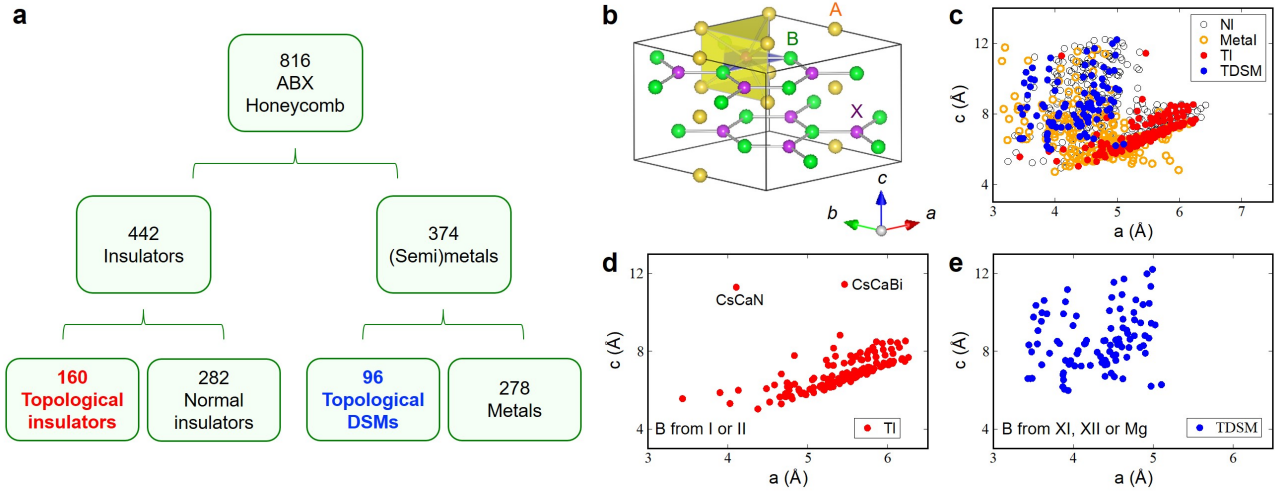

**Figure 1**. Classification of honeycomb ABX compounds in eight chemical groups: I-II-V, I-XII-V, XI-II-V, XI-XII-V, I-I-VI, I-XI-VI, XI-I-VI, and XI-XI-VI. (a) Sorting of 816 honeycombs ABX's into TI, NI, TDSM, and metal based on band structure and topological invariant calculations (see Section I in Supporting Information for detailed description of theoretical methods). (b) Crystal structure of honeycomb ABX compounds. The X-atom centered $XB_3A_6$ local motif is indicated by the yellow triangular prism and blue triangle. (c) Graphical separation of the ABX compounds into TI, NI, TDSM, and metal. (d-e), Separation of honeycomb ABX TI versus TDSM in (d) groups with B from I and II columns in the periodic table and (e) groups with B from XI and XII columns, or being Mg. The lattice constants $a$ and $c$ of the honeycomb structure are used as coordinates.

## II. Results and Discussions

### A. Search of topological materials from 816 ABX honeycomb compounds

Looking first for atomic sequences A/B/X at the fixed crystal structure we have performed high-throughput density functional theory (DFT) and topological invariant calculations on 816 real plus hypothetical honeycomb ABX compositions (**Figure 1**a), sorting them into topological (TI's or TDSM's) vs. trivial (normal insulators or metals). Our first-principles calculations show that the 816 honeycomb compounds are divided into 442 insulators plus 374 metals + semimetals. The latter are divided into 96 topological Dirac semimetals plus 278 ordinary metals, whereas the 442 insulators are divided into 160 TI's plus 282 normal insulators ($Z_2 = 0$ but could have non-symmorphic symmetry protected hourglass Fermion-like surface states such as in KHgSb[16]). We then identify within the first group of 256 topological compounds 160 TI's and 96 TDSM's by examining if the irreducible representations of the inverted bands are the same (implying band anti-crossing, hence TI) or different (implying band crossing, hence TDSM). The significant number of newly recognized TI's in this group as summarized in Figure 1a is noteworthy, but none overlap with those initially predicted to be TI based on a restricted view of the BZ.[12] Based on this classification, we have distilled a simple chemical regularity for the selectivity between TI and TDSM: The ABX compounds that are TDSM have B atoms (part of the BX honeycomb layers) that come from the Periodic Table columns XI (Cu, Ag, Au) or XII (Zn, Cd, Hg), or Mg (group II), whereas the ABX compounds whose B atoms comes from columns I (Li, Na, K, Rb, Cs) or II (Ca, Sr, Ba) are TI's.

### B. Separation of topological materials into TI's and TDSM's

We have performed DFT calculations with spin-orbit coupling (SOC) of electronic structures and topological invariants (see Section I in Supporting Information for the electronic structures[17-22] and topological invariants[9,23-25] evaluation methods) of the ABX honeycomb structures (see Figure 1). We are aware of the DFT errors on evaluating the band gaps of insulators and semiconductors that could affect the prediction of topological materials from DFT, and test the DFT results by comparing them with HSE[22] calculations (see Section II.B in Supporting Information showing that one out of four DFT predicted topological materials becomes normal insulator according to HSE calculations). On the other hand, the separation of topological materials into TI's and TDSM's, concerning the band-crossing/band-anti-crossing of band-inverted compounds, is not affected by the DFT errors on band gaps. We find that including SOC in structural relaxation has negligible effect on band-edge electronic states (see Section II.C in Supporting Information). Thus we perform structural relaxation without SOC, followed by electronic structure calculation with SOC that could increase band inversion energies in topological materials. In the honeycomb ABX lattice shown in Figure 1b the positions of the atoms are: T(A) = (0,0,0); T(B) = (1/3,2/3,1/4); T(X) = (2/3,1/3,1/4). Note that the B atom has as first and second shell neighbors $X_3A_6$, and the X atom has neighbors $B_3A_6$, whereas the A atom has as neighbors $[X_6B_6]A_6$ or $A_6[X_6B_6]$ (X and B are at equal-distance) depending on the lattice constants.

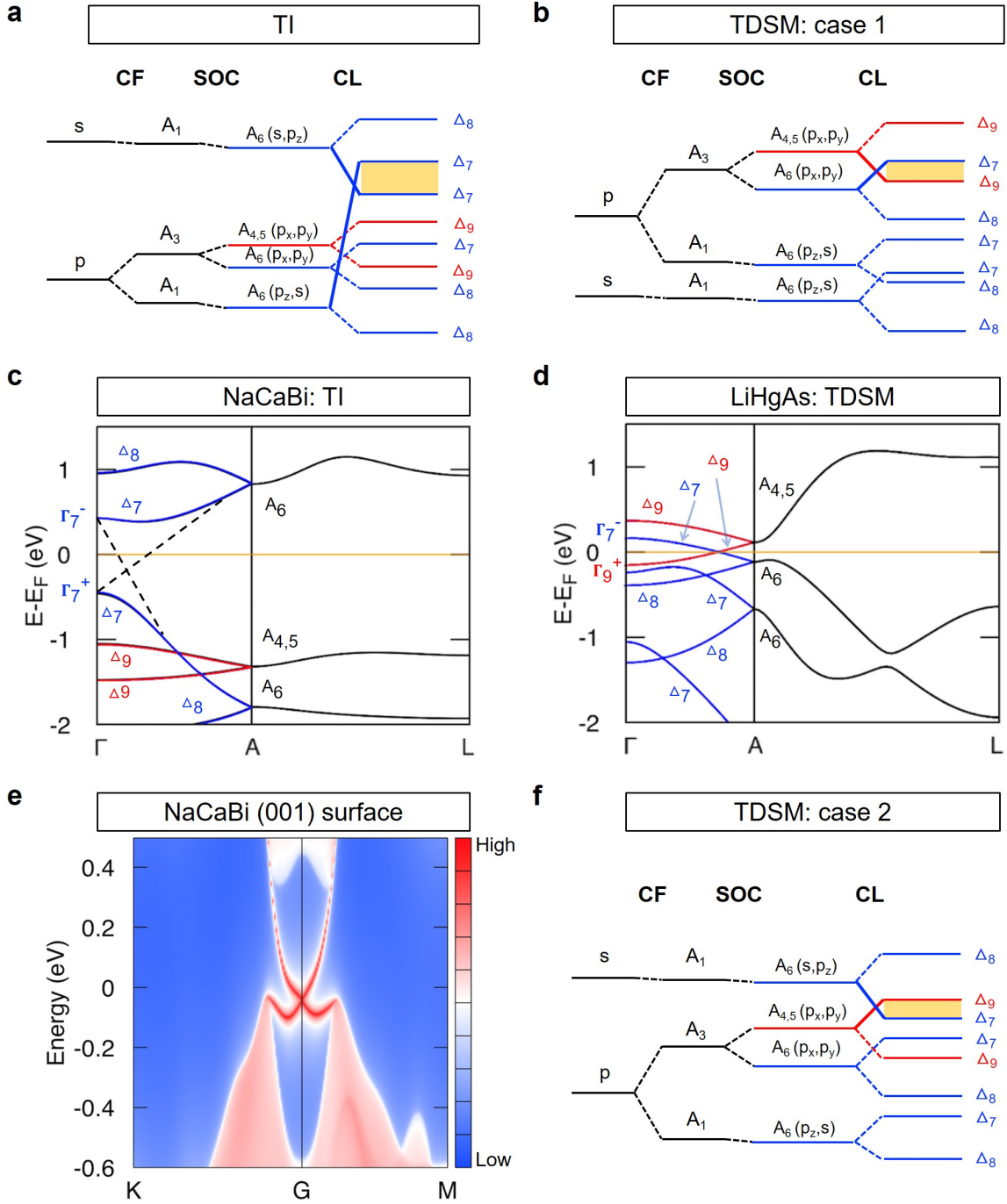

**Figure 2**. Electronic structures of TI and TDSM phases in honeycomb ABX structures. (a-b) Energy level diagrams for (a) TI with *s* above *p* state mainly from X atoms and (b) TDSM with *s* below *p* state showing the effect of crystal field (CF) splitting, SOC, and coupling between B-X layers (CL) on the states around the Fermi level. $A_1$ and $A_3$ ($A_{4,5}$ and $A_6$) are the irreducible representations at A point in the Brillouin zone in the absence (appearance) of SOC, and $\Delta_7$, $\Delta_8$, and $\Delta_9$ are those on the Γ-A line with SOC. The shaded areas indicate the inverted gap near the center of BZ. (c-d) Band structures of (c) NaCaBi (TI) and (d) LiHgAs (TDSM) with SOC. $\Gamma_7^+$, $\Gamma_7^-$, etc. are the irreducible representations at Γ point. Red and blue bands denotes $\Delta_9+\Delta_9$ and $\Delta_7+\Delta_8$ band character, respectively. We use $A_{4,5}$ as an abbreviation of $A_4+A_5$. The dashed lines in panel (c) are guidance to eyes for the band inversion. (e) (001) surface states of NaCaBi calculated by the effective tight-binding Hamiltonians of a freestanding thick slab consisting of 20 unit-cells along (001) direction. The color-scale indicates the relative density of surface states. (f) Energy level diagrams for TDSM with *s* above *p* state.

The topological invariants were calculated from band parities[26] for the centrosymmetric honeycomb ($P6_3/mmc$) structure, with structural parameters obtained by total energy and force minimization (see Section I in Supporting Information for theoretical methods). The classification of the 816 compounds into TI, TDSM, normal insulator (NI), and metals is given in Figure 1a and summarized graphically in terms of the lattice parameters *a* and *c* in Figure 1c-1e (see Figure S1 in Supporting Information for the classified list of the 816 compounds). We emphasize that although these structures are locally stable, there is no assurance that they are stable with respect to alternative structures or relative to phase separation into constituent phases. We nevertheless include all of them at this stage to clarify how the electronic feature of TI-ness or TDSM-ness emerges from a given composition at fixed structure. Filtering out stable structures will be discussed later for a subset of interesting compounds.

Figure 1 reveals a simple chemical regularity for the selectivity between TI and TDSM: The ABX compounds that are TDSM have B atoms (part of the BX honeycomb layers) that come from the Periodic Table columns XI (Cu, Ag, Au) or XII (Zn, Cd, Hg), or Mg (group II), whereas the ABX compounds whose B atoms comes from columns I (Li, Na, K, Rb, Cs) or II (Ca, Sr, Ba) are TI's. To understand this we will analyze next the origin of TI *vs.* TDSM in terms of symmetry (band crossing *vs.* band anti-crossing), and then connect these symmetries with geometrical factors such as *c/a* ratio controlling the crystal field splitting as well as the interaction between local structures, and hence crossing *vs.* anti-crossing.

**Band crossing (TDSM) vs. band anti-crossing (TI).** Both TI and TDSM phases have band inversion, but the group representations of the inverted bands could lead to either band crossing or anti-crossing off the TRIM point. This results in TDSM or TI, respectively. **Figure 2**a and 2b explains schematically the orbital conditions for TI and TDSM, respectively as further illustrated by actual band structure calculations for NaCaBi (a TI, Figure 2c) and LiHgAs (a TDSM, Figure 2d). This can be illustrated by considering the behavior along the $\Gamma$-A direction. With the presence of SOC, the bands at the A point are four-fold degenerate due to the non-symmorphic symmetry operation of space group P6$_3$/mmc.[27] There are two types of group representations at the wavevector A: $A_6$ and $A_4+A_5$ (see Table S2 in Supporting Information). In the direction from A to $\Delta$ (0, 0, $\Delta$) the four-fold degenerate states split into two twofold bands $A_6 \rightarrow \Delta_7+\Delta_8$ and $A_4+A_5 \rightarrow \Delta_9+\Delta_9$. Therefore, the bands at $\Delta$ have three possible double-group representations $\Delta_7$, $\Delta_8$ and $\Delta_9$. In the TI case as found in our calculations (Figure 2a), the inverted gap consists of the occupied $\Delta_7$ state, originating from $A_6$ and formerly a conduction (*s*) band, lying below the empty $\Delta_7$ state originating also from $A_6$, formerly a valence (*p*) band. When both inverted bands have the same $\Delta_7$ representation, there must be band anti-crossing and thus the system will be a TI, as shown in Figure 2a and exemplified by NaCaBi in Figure 2c. It is also shown in Figure 2c that the two inverted $\Delta_7$ states at $\Delta$ become at $\Gamma$ point $\Gamma_7^+$ and $\Gamma_7^-$, respectively, with opposite parities as indicated by plus/minus signs (see Table S3 in Supporting Information for their character table). The (001) surface states of NaCaBi with a clear Dirac cone is shown in Figure 2e, indicating a strong TI phase. In the TDSM case shown in Figure 2b and exemplified by LiHgAs in Figure 2d (indicating the parities of the inverted states $\Gamma_7^-$ and $\Gamma_9^+$ at $\Gamma$), the former conduction *s* band drops below the former valence *p*-band (see Figure 2b) due to the strong relativistic Mass-Darwin effect of Hg, hybridizing with the $p_z$ state, leaving the upper braches of the $p_{x,y}$ bands as conduction bands. The inverted gap of LiHgAs are between the occupied $\Delta_9$ state originating from $A_{4,5}$ (an abbreviation for $A_4+A_5$) and the empty $\Delta_7$ state originating from $A_6$ of the former valence *p*-band. Because these states originate from different ($A_{4,5}$ and $A_6$) symmetries, they can cross, leading to TDSM. Figure 2f shows another case of TDSM phase as found in our calculations (the band structure is exemplified by RbMgN in Figure S4 of Supporting Information), in which the conduction *s*-band does not drop below the valence *p*-band, analogous to the TI case (Figure 2a). However, in Figure 2f, the upper-most state of the valence *p*-band is the $\Delta_9$ state (*vs.* the $\Delta_7$ state in Figure 2a), and the inverted gap is between this $\Delta_9$ state and the $\Delta_7$ state originating from $A_6$ (former conduction *s*-band). The two states with different irreducible representations cross, leading to TDSM. Thus, the simple design principle for TI *vs.* TDSM classification of band-inverted compounds is whether the representations of the inverted bands are equal or different.

**The geometrical factors controlling crossing vs. anti-crossing and hence TI vs. TDSM in honeycomb structure.** Whereas the crossing *vs.* anti-crossing (Figure 2c, d) explains the *generic* TDSM *vs.* TI behavior, the understanding of *which compounds and lattice constants* fall into TDSM *vs.* TI behavior requires a more detailed understanding of crystal field splitting, SOC, and coupling between B-X layers. This is explained in Figure 2a, 2b, and 2f, for the 3 types of TI/TDSM electronic structures found in our calculations. For the appearance of TI case (Figure 2a), a $p_z$-like $\Delta_7$ state needs to be pushed to the uppermost state of the valence *p*-band, thus a small *c/a* ratio is preferred for raising the $p_z$ states up relative to the $p_{x,y}$ states. Furthermore, strong coupling between B-X layers conducted by intercalated A atoms can split the $p_z$ states into two branches, pushing one branch up. The coupling between B-X layers is not solely determined by the *c* lattice constant, but also by the *c/a* ratio, since A atom has equal-distance nearest B and X neighbors only for small *c/a* ratio. Therefore, the predicted TI's are mainly distributed in the region with small *c* and small *c/a* ratio (see Figure 1c, 1d), with only two exceptions (CsCaN and CsCaBi) that have very large A atom (Cs) and thus expanded *c* lattice constant (see Figure S5 in Supporting Information for their band structures).

It is interesting to find that all ABX TI's (TDSM's) have the B atom from columns I (Na) and II (Ca) [XI (Cu) and XII (Zn)] in the periodic table with electronic configuration *[...]s$^n$* (*[...]d$^{10}$s$^n$*), with the exceptions for B = Mg (group II) (see their band structures in Figure S4 of Supporting Information). These trends and exceptions are quite understandable from the view of elemental sizes since Mg has similar atomic/ionic sizes as the B from XII and XI but much smaller atomic/ionic sizes than the other II and I species (Ca, Sr, Ba, Li, Na, K, Rb, Cs). Furthermore, the strong hybridization between B-*d* and X-*p* valence states for B elements with *[...]d$^{10}$s$^n$* electronic shells (see e.g. Figure S6 in Supporting Information for the case of LiAuTe) could decrease the B-X distance, leading to large *c/a* ratio and smaller coupling between B-X layers thus preference for TDSM.

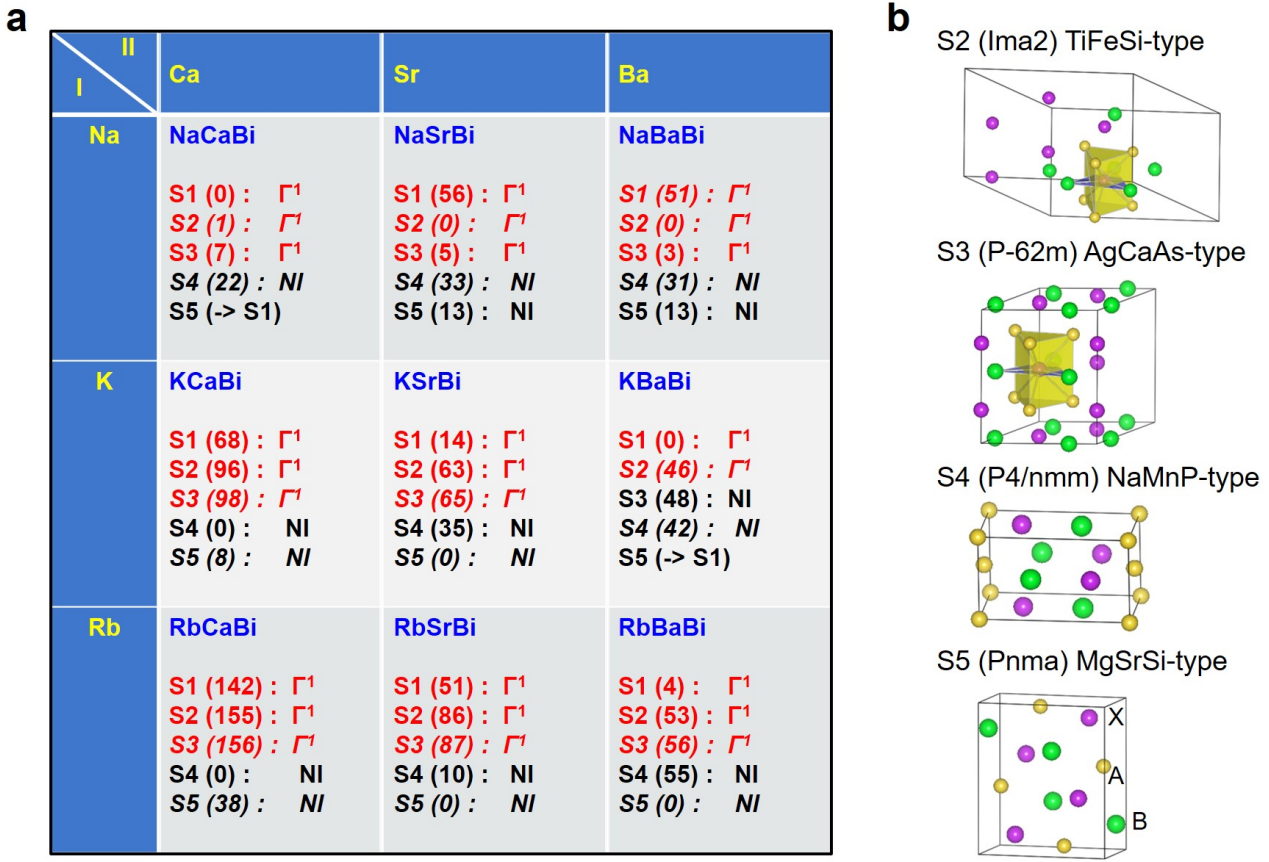

**Figure 3.** Topological properties of the I-II-Bi group of ABX materials. (a) Topological (red) and non-topological (black) phases in I-II-Bi with the crystal structures observed[31] for the compounds in this group (denoted S1-S5). Italics font indicates BAX configuration with cation swapping comparing to ABX (we choose the lower-energy configuration from ABX and BAX). The total energies relative to the ground state structure are given in parentheses (in meV/atom). (b) Crystal structures of the I-II-Bi group of materials. The honeycomb structure (S1) is shown in Figure 1b.

**C. Co-evaluation of stability and TI-ness: The ABX Bismides as novel TI's**

***Stable ABX compounds that are TI's.*** We will next focus on ABX compounds that are both TI and stable. This inquiry is motivated by the recent recognition[28] that band inversion—i.e. population of anti-bonding conduction states and depopulation of bonding valence states—while promoting topological character, can also thermodynamically destabilize the compound if it such inversion occurs in a significant fraction of the Brillouin zone in compounds unable to strongly screen the ensuing atomic displacements (such as ionic oxides[28]). Thus one must examine if the predicted structures are stable dynamically[29,30] and weather a structure is stable against decomposition into competing phases[31] (see Figure S7 in Supporting Information for the case of NaCaBi).

The compounds and structures indicated in **Figure 3** in red are predicted TI whereas black indicates normal insulators. Examination of the above noted thermodynamic stability, we identify two ABX compounds NaCaBi and KBaBi that are stable in the honeycomb structure. Furthermore, we identify another 7 compounds I-II-Bi (with I = Na, K Rb; II = Ca, Sr, Ba) that are thermodynamically stable in other structures (denoted S2-S5 in Figure 3 with S1 being honeycomb structure) against disproportionation in their competing phases. In addition to the honeycomb-stable TI's NaCaBi and KBaBi, we find another two compounds NaSrBi and NaBaBi are stable in S2 structure and predicted to be TI's from DFT+SOC (NaBaBi in a slightly higher-energy structure S3 was predicted to be TI before [32]). **Figure 4** shows the electronic structures of the predicted TI's in the lowest-energy structures with blue arrows indicating their band inversions. Their (inversion energies, fundamental band gaps) are (0.89, 0.34) eV for NaCaBi (S1), (0.33, 0.22) eV for KBaBi (S1), (0.20, 0.05) eV for NaSrBi (S2), and (0.05, 0.05) eV for NaBaBi (S2). In addition to the $k$-path adopted in Fig. 4a,b and Figs. S2 and S3 in Supporting Information for honeycomb TI's and TDSM's, we checked another $k$-path (Γ-M-K-Γ-A-L-H-A) for all the 816 ABX honeycombs (see e.g. Figs. S4 and S5 in Supporting Information). The above two $k$-paths include all the high symmetry $k$-lines in the Brillouin zone of honeycomb (S1) structure. In the predicted honeycomb topological materials, we didn't find any Dirac points away from Γ-A $k$-line where the $C_3$ rotation symmetry can protect the 4-fold Dirac point. Since all the predicted TDSM's are in the centrosymmetric honeycomb structure, the 2-fold Weyl nodes (band crossing by two singly degenerated bands at arbitrary $k$-points) do not exist. Figure 3 also shows that all the ABX compounds stable in S4 and S5 structures are normal insulators. We next attempt to understand this trend based on the structural features of this sub-group of ABX compounds.

**The TI-enabling structural motif for stable ABX Bismide TI.** The S1 (honeycomb), S2 and S3 structures that are TI's have a similar structure motif $BiA_6B_3$, i.e. Bi surrounded by an $A_6$ triangular prism and intercalated with a $B_3$ triangle, as shown in Figure 1 and Figure 3, whereas this motif is missing in S4 and S5 structures that are stable but not TI's. We have calculated the properties of these five structures considering ABX and permutated BAX configurations for the 9 compounds with DFT+SOC, and systematically find the same situation as above, i.e. all structures without $BiA_6B_3$ motif (S4, S5) are normal insulators, whereas all structures with the $BiA_6B_3$ motif (S1, S2, S3) are TI's. There is but a very weak exception (KBaBi in S3 structure with a band gap of 14 meV from DFT+SOC, see Figure S8 in Supporting Information).

We have seen that the size factor (Figure 1) as represented by $a$, $c/a$ in honeycomb structure, especially the size of the B atom (e.g. Mg and Zn versus Ca and Sr), is related the to the selectivity within the ABX honeycomb lattice to TI versus TDSM in band-inverted ABX. We further note that the splitting of the $p_z$ states due to inter-layer coupling can push one of $p_z$ state up to VBM. The underline factor is the anisotropic size of the local motifs ($XA_6B_3$, see Figure 1b and Figure 3b) in the periodic structure that can also affect the band inversion, e.g. if the inter-layer coupling is reduced, the VBM $p_z$ state moves down, then the band inversion could be removed. The CBM $s$ state will also be affected by the size of the local motif. The high coordination number of Bi surrounded by positively charged ions in the large-size $BiA_6B_3$ motif helps delocalize CBM Bi-$s$ state towards cations and shift the Bi-$s$ level down (the relativistic Mass-Darwin effects also shift down the Bi-$s$ level). Furthermore, highly symmetrized structure of the $XA_6B_3$ motif increases state repulsions shifting VBM (CBM) up (down) and thus helps form band inversions, analogous to the highly symmetrical $BO_6$ motif in $ABO_3$ TI's.[28] In Figure 3, we see a TI to narrow-gap (~14 meV) normal insulator transition along with the cation substitution from KCaBi to KBaBi, which could be related to the variation of the motif sizes and shapes due to cation variations.

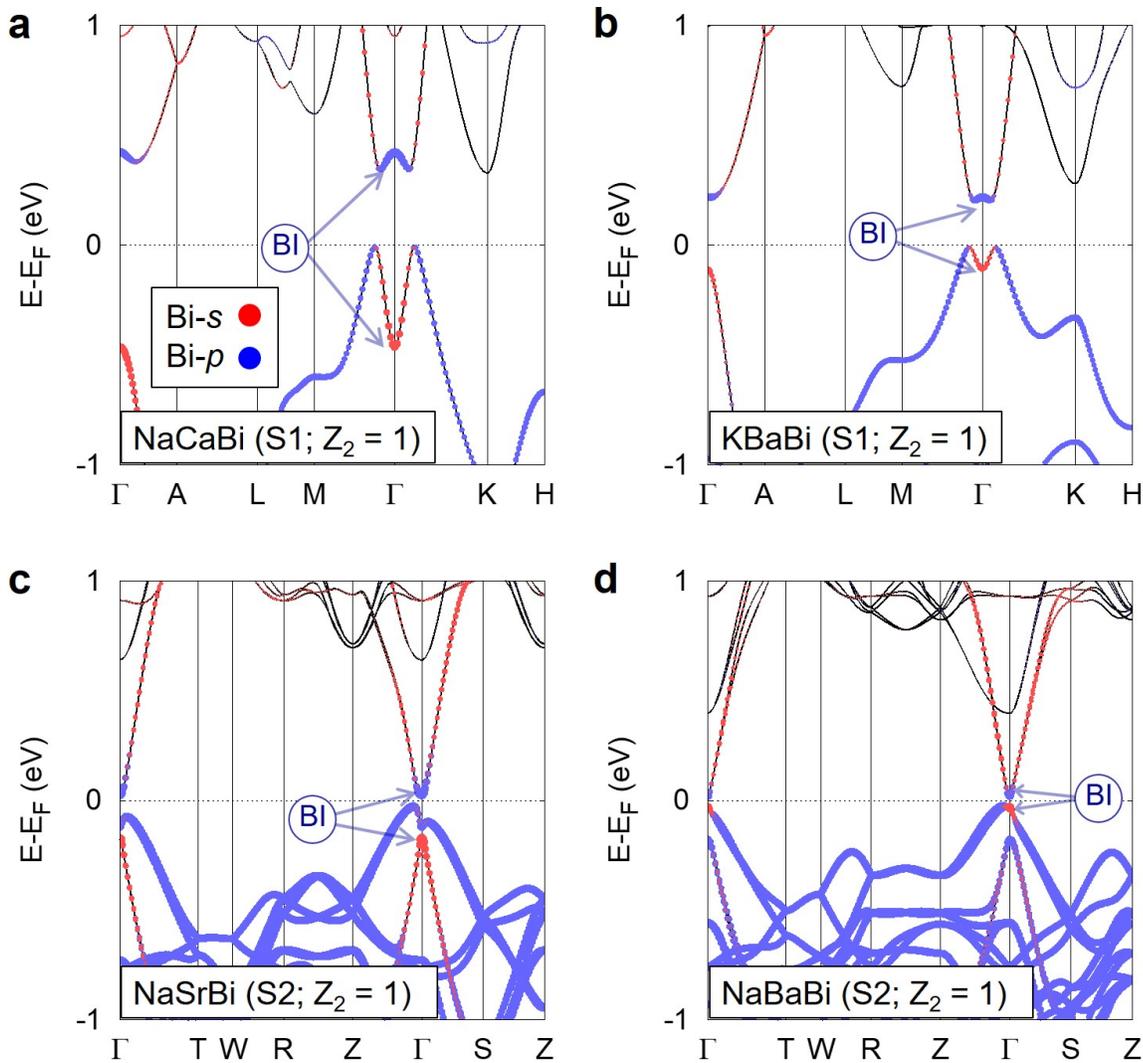

**Figure 4.** Electronic structures of the predicted TI's in the lowest-energy structures from DFT+SOC. (a) Orbital-projected band structure of NaCaBi in the honeycomb (S1) structure (see Figure S2 in Supporting Information for the band-inverted electronic structure of NaCaBi from HSE06 [22]); (b) KBaBi (S1); (c) NaSrBi in the S2 structure (see Figure 3b); (d) NaBaBi (S2). The dotted lines with different colors denote the band projection onto different atomic orbitals. The band inversion is denoted in the figure by BI, with arrows pointing to the inverted states.

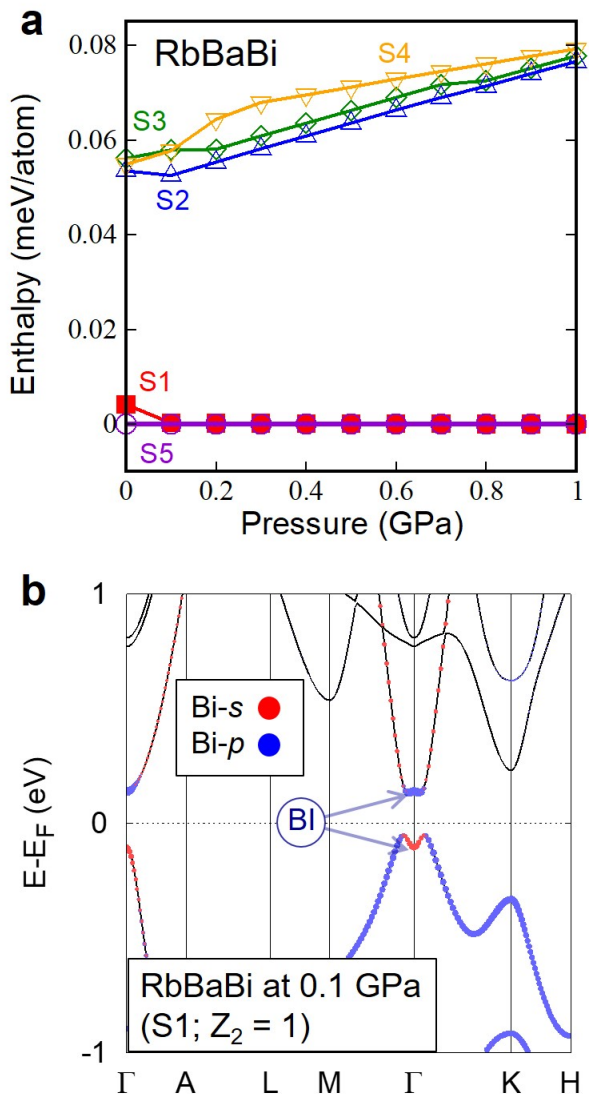

**Figure 5.** Pressure stabilized TI phase in RbBaBi. (a) Enthalpy of crystal structures (shown in Figure 1 and Figure 3) of RbBaBi under hydrostatic pressure. We choose the lower-energy from ABX and BAX (for cation swapping) configurations. (b) Electronic structure of honeycomb (S1) RbBaBi under pressure of 0.1 GPa from DFT+SOC. The dotted lines with different colors denote the band projection onto different atomic orbitals. The band inversion is denoted by BI, with arrows pointing to the inverted states.

***Pressure stabilized honeycomb Bismide TI.*** Similar to the $BiA_6B_3$ motifs discussed above, in Ref. [28], a $BO_6$ motif is found to be responsible for the band inversion in $ABO_3$ TI's. Furthermore, it is found that large enough distortions to the $BO_6$ motif can remove the band inversion, and once the distortions are *partially* or fully restored (e.g. by external pressure), the band inversion could be recovered.[28] Many of ABX compounds predicted to be TI's in the honeycomb structure in Figure 3a are stable in non-honeycomb structures at ambient conditions (similarly for the many topological materials in Figure 1c), such as RbBaBi with lowest-energy structure S5 (Pnma, see Figure 3b), which is an anti-ferroelectrically dis-torted version of S1 (honeycomb) structure.[33,34] We consider the possibility of restoring the anti-ferroelectric distortion and thus stabilizing the honeycomb TI structures by pressure. As an example, we apply hydrostatic pressure to RbBaBi that is a normal insulator at ambient condition in its ground state structure (Pnma, see Figure 3b), and find that a 0.1 GPa pressure on RbBaBi can induce a transition from the Pnma structure to the honeycomb structure, as shown in **Figure 5**a. Figure 5b shows that the stable honeycomb RbBaBi at 0.1 GPa is a TI, indicating that a rather small decrease of lattice volume by the external pressure (comparable to the variation of relaxed volumes from generalized gradient approximation (GGA) to, local density approximation (LDA)[35,36] to DFT as applied in Ref. [34]), can remove the anti-ferroelectric distortion and stabilize the TI phase. This result suggests that once the TI-enabling structural motif is identified, one can design external constraints to achieve such structural motif and the related TI phase.

### III. Conclusions

We found from DFT calculations 256 new topological materials in the family of ABX honeycomb compounds, and distilled an interesting chemical rule to separate them into TI's and TDSM's simply by the identity of the B species in the formula. The chemical rule is based on the analysis of the double group representations of the band edge states in the honeycomb ($P6_3/mmc$) lattice that are determined by the interplay between ionic sizes and atomic orbitals in the local structures (i.e. $X[B_3A_6]$) associated with topological properties. The simple chemical regularity shed light on the high-throughput design of complicated and subtle materials functionalities (e.g. the TDSM *vs*. TI that requires detailed analysis of irreducible representations of the states near Fermi level), i.e. instead of performing elaborated study of each structure, the chemical rule distilled from a set of materials can be used to guide fast assignment of functional materials based on their chemical formulae, followed by verification on representative cases.

The special local motif, $X[B_3A_6]$, was found to coincide with the appearance of topological band inversion in various structure types of the I-II-Bi group of materials, i.e. structures with (without) such motif are TI (NI) with a negligible exception case. For a structure without such motif and thus not TI, we applied external pressure to reinforce the needed structural motif and achieved a TI phase at pressure of 0.1 GPa. The functionality-enabling structural motif and the possibility to realize it in materials where it does not appear naturally, open the way of target-oriented design of functional materials, i.e. instead of finding a special functionality from the variety of structure motifs by accident, we can first identify the functionality-enabling structural motifs via rational study of structures that are not necessarily stable at ambient conditions, then using these special motifs to construct potential functional materials, followed by stability test or design of external constraints to stabilize the designed materials.

### ASSOCIATED CONTENT

#### Supporting Information

The Supporting Information is available free of charge on the ACS Publications website.

Full list of studied ABX honeycomb structures, theoretical methods, additional figures and tables on the properties of ABX topological materials (PDF)

## AUTHOR INFORMATION

### Corresponding Author
*E-mail: alex.zunger@colorado.edu. Phone: 1-303-492-7084.
*E-mail: xiuwenzhang@szu.edu.cn. Phone: 86-755-2690-1883.

### Author Contributions
‡X.Z., Q.L., and Q.X. contributed equally to this work. / All authors have given approval to the final version of the manuscript.

### Funding Sources
Department of Energy, Office of Science, Basic Energy Science, MSE division.
### Notes
The authors declare no competing financial interest.

## ACKNOWLEDGMENT

The work of A. Z. and X. Z. was supported by Department of Energy, Office of Science, Basic Energy Science, MSE division under Grant No. DE-FG02-13ER46959 to CU Boulder. This research used resources of the National Energy Research Scientific Computing Center (NERSC), a U.S. Department of Energy Office of Science User Facility operated under Contract No. DE-AC02-05CH11231. The work of X. Z. in China was supported by National Natural Science Foundations of China (Grant No. 11774239), National Key R&D Program of China (Grant No. 2016YFB0700700), and Shenzhen Science and Technology Innovation Commission (Grant No. JCYJ20170412110137562, JCYJ20170818093035338).
## REFERENCES

(1) Hasan, M. Z.; Kane, C. L. *Reviews of Modern Physics* **2010**, *82*, 3045.
(2) Qi, X.-L.; Zhang, S.-C. *Rev. Mod. Phys.* **2011**, *83*, 1057.
(3) Kane, C. L.; Mele, E. J. *Phys. Rev. Lett.* **2005**, *95*, 146802.
(4) Bernevig, B. A.; Hughes, T. L.; Zhang, S.-C. *Science* **2006**, *314*, 1757.
(5) Wang, Z.; Sun, Y.; Chen, X.-Q.; Franchini, C.; Xu, G.; Weng, H.; Dai, X.; Fang, Z. *Phys. Rev. B* **2012**, *85*, 195320.
(6) Liu, Z. K.; Zhou, B.; Zhang, Y.; Wang, Z. J.; Weng, H. M.; Prabhakaran, D.; Mo, S.-K.; Shen, Z. X.; Fang, Z.; Dai, X.; Hussain, Z.; Chen, Y. L. *Science* **2014**, *343*, 864.
(7) Zhang, H. J.; Liu, C.-X.; Qi, X.-L.; Dai, X.; Fang, Z.; Zhang, S.-C. *Nature Physics* **2009**, *5*, 438.
(8) Gu, Z.-C.; Wen, X.-G. *Phys. Rev. B* **2009**, *80*, 155131.
(9) Yang, B.-J.; Nagaosa, N. *Nature Communications* **2014**, *5*, 4898.
(10) Young, S. M.; Zaheer, S.; Teo, J. C. Y.; Kane, C. L.; Mele, E. J.; Rappe, A. M. *Phys. Rev. Lett.* **2012**, *108*, 140405.
(11) Liang, T.; Gibson, Q.; Ali, M. N.; Liu, M.; Cava, R. J.; Ong, N. P. *Nature Materials* **2015**, *14*, 280.
(12) Zhang, H. J.; Chadov, S.; Müchler, L.; Yan, B.; Qi, X. L.; Kübler, J.; Zhang, S. C.; Felser, C. *Phys. Rev. Lett.* **2011**, *106*, 156402.
(13) Gibson, Q. D.; Schoop, L. M.; Muechler, L.; Xie, L. S.; Hirschberger, M.; Ong, N. P.; Car, R.; Cava, R. J. *Phys. Rev. B* **2015**, *91*, 205128.
(14) Du, Y.; Wan, B.; Wang, D.; Sheng, L.; Duan, C.-G.; Wan, X. *Sci. Rep.* **2015**, *5*, 14423.
(15) Zhang, X.; Liu, Q.; Luo, J. W.; Freeman, A. J.; Zunger, A. *Nat. Phys.* **2014**, *10*, 387.
(16) Wang, Z.; Alexandradinata, A.; Cava, R. J.; Bernevig, B. A. *Nature* **2016**, *532*, 189.
(17) Kohn, W.; Sham, L. J. *Phys. Rev. A* **1965**, *140*, 1133.
(18) Kresse, G.; Joubert, D. *Phys. Rev. B* **1999**, *59*, 1758.
(19) Perdew, J. P.; Burke, K.; Ernzerhof, M. *Phys. Rev. Lett.* **1996**, *77*, 3865.
(20) Kresse, G.; Furthmüller, J. *Comput. Mater. Sci.* **1996**, *6*, 15.
(21) Heyd, J.; Scuseria, G. E.; Ernzerhof, M. *J. Chem. Phys.* **2003**, *118*, 8207.
(22) Heyd, J.; Scuseria, G. E.; Ernzerhof, M. *J. Chem. Phys.* **2006**, *124*, 219906.
(23) Fu, L.; Kane, C. L.; Mele, E. J. *Phys. Rev. Lett.* **2007**, *98*, 106803.
(24) Fu, L.; Kane, C. L. *Phys. Rev. B* **2006**, *74*, 195312.
(25) Yu, R.; Qi, X. L.; Bernevig, A.; Fang, Z.; Dai, X. *Phys. Rev. B* **2011**, *84*, 075119.
(26) Fu, L.; Kane, C. L. *Phys. Rev. B* **2007**, *76*, 045302.
(27) Liu, Q.; Zunger, A. *Phys. Rev. X* **2017**, *7*, 021019.
(28) Zhang, X.; Abdalla, L. B.; Liu, Q.; Zunger, A. *Adv. Funct. Mater.* **2017**, *27*, 1701266.
(29) Gautier, R.; Zhang, X.; Hu, L.; Yu, L.; Lin, Y.; Sunde, T. O. L.; Chon, D.; Poeppelmeier, K. R.; Zunger, A. *Nature Chemistry* **2015**, *7*, 308.
(30) Zhang, X.; Yu, L.; Zakutayev, A.; Zunger, A. *Adv. Funct. Mater.* **2012**, *22*, 1425.
(31) *Inorganic Crystal Structure Database, Fachinformationszentrum Karlsruhe, Germany, (2006)*.
(32) Sun, Y.; Wang, Q.-Z.; Wu, S.-C.; Felser, C.; Liu, C.-X.; Yan, B. *Phy. Rev. B* **2016**, *93*, 205303.
(33) Bennett, J. W.; Garrity, K. F.; Rabe, K. M.; Vanderbilt, D. *Phys. Rev. Lett.* **2013**, *110*, 017603.
(34) Monserrat, B.; Bennett, J. W.; Rabe, K. M.; Vanderbilt, D. *Phys. Rev. Lett.* **2017**, *119*, 036802.
(35) Zunger, A.; Perdew, J. P.; Oliver, G. L. *Solid State Commun.* **1980**, *34*, 933.
(36) Perdew, J. P.; Zunger, A. *Phys. Rev. B* **1981**, *23*, 5048.

For Table of Contents Graphic Only

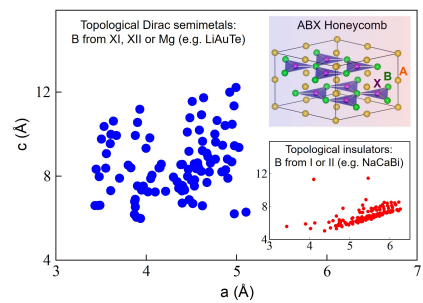